\documentclass[referee]{aa}       % for a referee version
\usepackage{graphics}
\usepackage{epsfig}
\usepackage{latexsym}
\def\E$\gamma${E_$\gamma$}

\def \deg      {$^{\circ}$}

\def \sig      {$\sigma$}
\def \gray     {$\gamma$-ray}
\def \grays    {$\gamma$-rays}

\begin{document}

%\thesaurus{   
%               03                    % Main Journal (?) 
%              (13.07.2;              % Gamma rays: observations
%               11.01.2;              % Galaxies: active
%               11.17.4 3C~273);      % Galaxies: quasars: individual: 3C273
%              }
%

\title{COMPTEL Observations of the $\gamma$-ray Blazars 3C~454.3 and CTA~102 During the CGRO Mission}

\author{S.~Zhang\inst{1,2},
        W.~Collmar\inst{2}, V.~Sch\"onfelder\inst{2}
}
 
\institute{Laboratory for Particle Astrophysics, 
              Institute of High Energy Physics,
             P.O.Box 918-3, Beijing 100049, China
             \and
             Max-Planck-Institut f\"ur extraterrestrische Physik,
               Postfach 1613, 85741 Garching, Germany 
          }

\offprints{S.~Zhang}
\mail{szhang@mail.ihep.ac.cn}

\date{Received  / Accepted }

\titlerunning{COMPTEL Observations of the \gray\ Blazars 3C 454.3 ...}
\authorrunning{S.~Zhang et al.}

\abstract{
We have investigated the MeV behaviour of the \gray\ blazars 3C~454.3 and 
CTA~102 by analyzing all COMPTEL observations of this sky region 
during the complete CGRO mission.
Both sources are detected by COMPTEL at the upper COMPTEL energies; 
although their flux estimates may be uncertain by possible minor
contribution of nearby unidentified EGRET \gray\ sources.
While CTA~102 was only detected at energies above 10~MeV during 
the early mission, 3C~454.3 is most significantly detected in the 
COMPTEL 3-10~MeV band in the sum of all data. Time-resolved analyses 
indicate a weak (near COMPTEL threshold) but likely steady 3-10~MeV
emission over years, being independent of the observed time variability 
at energies above 100~MeV as observed by EGRET. This energy-dependent 
variability behaviour suggests different emission mechanisms at work 
at the two bands. 
Putting the COMPTEL fluxes in multifrequency perspective (radio to \grays),
reveals for both sources the typical two-hump blazar spectra,
with a low-energy maximum around the IR and a high-energy maximum 
at MeV energies. The latter one dominates the energy output across
the whole electro-magnetic spectrum. The results of our analyses are discussed
in the framework of current blazar modeling.  

}

\maketitle

\section{Introduction}
The Compton Gamma-Ray Observatory (CGRO), carrying four individual 
\gray\ experiments (EGRET, COMPTEL, OSSE, BATSE), explored 
the \gray\ sky for more than 9 years
between April 1991 and June 2000. Due to the improved sensitivities,
the wide field-of-view of the instruments, and the long mission/exposure time,
many new \gray\ sources were detected.  For example, the EGRET telescope,
sensitive to 
photon energies between 30~MeV and $\sim$10~GeV, detected $\sim$90 
blazar-type Active Galactic Nuclei (AGN) (Hartman et al. 1999). 
These sources are characterized by a compact and radio-loud core,
a flat radio spectrum (S$\propto\nu^{\alpha}$, $\alpha$$>$-0.5),
and high flux variability at all energy bands.
Among them are the quasars 3C~454.3 and CTA~102, which were discovered 
to be \gray\ emitters during the early CGRO mission in 1991 and 1992
(Nolan et al. 1993, Hartman et al. 1993).
Both blazars were also detected at soft \grays\ and hard X-rays 
by the CGRO experiments COMPTEL (0.75-30~MeV) and OSSE (0.05-10~MeV).
3C~454.3 (l/b:86.11\deg/-38.18\deg, $z = 0.859$) and
CTA~102 (l/b: 77.44\deg/-38.58\deg, $z = 1.037$) are prominent radio 
sources showing both superluminal motion
(Unwin et al. 1989, Baath 1987).

Both sources belong to the OVV (optically violently variable) class
(Angel $\&$ Stockman 1980) and the HPQ (high polarization quasars)
class of quasars. The optical polarization can be as high as 11$\%$
for CTA~102 (Moore $\&$ Stockman 1981) and is on average 3$\%$
for 3C~454.3 (Wills et al. 1992).

At soft X-ray energies the sources show also the typical blazar signs,
i.e. flux variability and a power-law shaped spectrum of energy index
$\sim$0.6 for 3C~454.3 (Comastri et al. 1997). 
At hard X-rays, 3C~454.3 and CTA~102 were occasionally detected by OSSE 
at energies between 50 keV and 1~MeV, both showing flux variability
(McNaron-Brown et al. 1995).

Blom et al. (1995) reported the COMPTEL MeV-detections of both sources
during the early CGRO observations in 1991 and 1992.
According to the third EGRET catalog (Hartman et al. 1999) both blazars
are repeatedly detected by EGRET at energies above 100~MeV, though 3C~454.3
on average at a higher flux level than CTA~102.  

To generally investigate the MeV properties of both sources, 
we have consistently analyzed/reanalyzed all COMPTEL data, 
applying the most recent analyses methods. 
This supersedes the work of Blom et al. (1995), who reported the 
COMPTEL results on 3C~454.3 and CTA~102 from observations in 1991 and 1992. 
The analyses were carried out by first reanalyzing the early mission 
data, and then extending to the complete COMPTEL data base.
We describe the instrument and the data analysis procedure in Sect.~2,
the CGRO observations in Sect.~3, give the results and their discussion
in Sect.~4 and 5, and finally conclude in Sect.~6.

%######################################################################
\section{Instrument and data analysis}
%######################################################################
The imaging Compton Telescope COMPTEL was sensitive to $\gamma$-rays in 
the energy range 0.75-30 MeV with an energy-dependent energy
and angular resolution of 5$\%$ - 8$\%$ (FWHM) and 
1.7\deg - 4.4\deg (FWHM), respectively. 
It had a large field of view of about 1 steradian and was able
to detect $\gamma$-ray source with an location accuracy of the order
of 1$^{\circ}$-2$^{\circ}$, depending on source flux. For the 
details about COMPTEL see Sch\"onfelder et al. (1993).

COMPTEL contained two detector arrays in which an incident $\gamma$-ray
photon is first Compton scattered in a detector of the upper detector array
and -- in the favorable case -- then interacts with a detector of the
lower detector array. The scattered photon direction ($\chi$,$\psi$)
is obtained from the interaction locations in the two detectors. 
The Compton scatter angle $\bar{\varphi}$ is calculated from the
measured energy deposits of the photon in the two detectors. 
These quantities, scatter direction and angle, constitute a
three-dimensional data space in which the spatial response of
the instrument is cone-shaped and standard imaging methods, e.g. 
maximum entropy and maximum likelihood, are applied.
In the COMPTEL data analysis package the maximum-likelihood method
is used to estimate source parameters like detection 
significances, fluxes and flux errors. 
The detection significance is calculated from the 
quantity -2ln$\lambda$, where $\lambda$ is the ratio of the
likelihood L$_{0}$ (background) and the likelihood L$_{1}$
(source + background). The quantity -2ln$\lambda$ has
a $\chi_{3}^{2}$
distribution (3 degrees of freedom) 
for a unknown source and a $\chi_{1}^{2}$ distribution for 
a known source (de Boer et al. 1992).
The instrumental COMPTEL background was modelled by the 
standard filter technique in data space (Bloemen et al. 1994).
Due to the high galactic-latitude location of the sources,
no diffuse emission models were included in the analysis.
To estimate the source flux, we applied 
instrumental point spread functions assuming an E$^{-2}$ power
law shape for the source input spectrum.
We note that the derived fluxes are weakly
dependent on this particular shape. 
3C~454.3 and CTA~102 are neighboring sources within the COMPTEL
field-of-view and therefore their fluxes were estimated by  
simultaneous flux fits.

%######################################################################
\section{Observations}
%######################################################################
During the complete CGRO mission from April 1991 to June 2000,
in 14 observational periods -- so-called CGRO viewing periods (VPs),
each lasting for typically 1 to 2 weeks --  
the offset angles (angle between the source position and the
pointing direction of
COMPTEL) were less than 30 degrees for either 3C~454.3 or CTA~102.
Table~1 shows, that these 14 VPs are grouped 
within 6 so-called CGRO Phases. 
A CGRO Phase covers a time period of roughly 1 year. The whole 
CGRO mission was subdivided into 9 Phases, and COMPTEL observations
on the two blazars are available  up to Phase 7 in 1998. 
These 14 VPs add up to total effective exposures (100\% COMPTEL 
pointed directly to the source) of 27.19 days for 3C~454.3 and 27.32 days 
for CTA~102.

\section{Results}
%######################################################################
\subsection{Reanalysis of early-mission data}
During the COMPTEL mission, the data analysis methods (e.g. background modeling, 
event selections) have been improved compared to the earlier Phases. 
We reanalyzed the early-mission data,
published by Blom et al. (1995), using the most up-to-date methods. 
Our results are generally consistent with those published by Blom et al. (1995).
We find a significant excess in the uppermost (10 - 30~MeV) COMPTEL band,
which is consistent with the locations of 3C~454.3 and CTA~102 (Fig.~1). 
However, EGRET has detected three more \gray\ sources in this sky region
(Hartman et al. 1999): 3EG~J2255+1943 
($\alpha,\delta$ = 343.99$^{\circ}$,19.73$^{\circ}$; 
tentatively identified with the blazar PKS~2250+1926), 
3EG~J2243+1509 ($\alpha,\delta$ = 
340.78$^{\circ}$,15.17$^{\circ}$; unidentified), and 3EG~J2248+1745 
($\alpha,\delta$ = 342.24$^{\circ}$,17.77$^{\circ}$; unidentified).
They are also positionally consistent with this MeV emission (Fig.~1). 
Due to the poorer angular resolution of COMPTEL compared to EGRET,
these five known EGRET sources cannot be resolved by COMPTEL. 
In order to elaborate on this MeV excess, we considered the time 
behaviour of these 5 sources as observed by EGRET.  
According to the third EGRET catalog (Hartman et al. 1999), 
3EG~2255+1943 was not  
detected in any of these early-mission VPs,
and therefore we consider its contribution to this 10-30 MeV excess as negligible.
3EG~J2243+1509 was only detected in VP 26
($\sim$4$\sigma$ level), and 3EG~J2248+1745 only
in VP 28 ($\sim$3$\sigma$ level).
In contrast, 3C~454.3 and CTA~102 were detected in all of these early 
VPs. For the sum of the four VPs in Phase 1,
the third EGRET catalog lists the following
fluxes (units 10$^{-8}cm^{-2}s^{-1}$): 75.0$\pm$6.8 for 3C~454.3,
27.7$\pm$4.5 for CTA~102, $<$9.9 for 3EG~2255+1943, 
9.7$\pm$4.7 for 3EG~J2243+1509, and $<$20.8 for 3EG~J2248+1745. 
The extrapolations of the averaged spectra to MeV energies show
that 3C~454.3 and CTA~102 should have the strongest MeV emission (Fig.~1).
Therefore we assume that the majority of the detected 
MeV emission should be due to 3C~454.3 and CTA~102, because they are 
the strongest EGRET sources during this time period.

To test our assumption, we first analyzed the four VPs individually.
The results are not conclusive because in individual VPs
the MeV emission is too weak to be attributed to individual sources.
Then we analyzed separately the combinations of VPs 19, 28, 37
(3EG~2243+1509 not detected by EGRET), VPs 19, 26, 37 (3EG~2248+1745
not detected by EGRET), and VP 19, 37 (only 3C~454.3 and CTA~102 detected
by EGRET).
In all cases an MeV excess remains, though accordingly reduced in significance.
We like to note that the 3C~454.3 and CTA~102 fluxes, obtained
from the sum of the 4 VPs (Table~2), are consistent with those
from only VPs 19+37 within error bars.  
And finally, analyzing the sum of the four VPs by assuming 
3C~454.3 and CTA~102 being the only emitting sources and removing their
contributions from the map, provides a rather 
empty COMPTEL map with some remaining emission from the location of
3EG~J2243+1509.

In summary, the excess is consistent with the main contributions 
being from 3C~454.3 and CTA~102, although this cannot be finally proven 
due to the insufficient angular resolution of COMPTEL.
Their flux estimate is uncertain by the probable minor contributions
of the two unidentified sources in VP~26 and 28.

The simultaneous COMPTEL/EGRET spectra of 3C~454.3 and CTA~102 of
VP~19 are shown in Fig.~2. The EGRET spectral shapes 
are from Hartman et al. (1993) for 3C~454.3 and from 
Nolan et al. (1993) for CTA~102.
The COMPTEL spectral data are consistent with the extrapolations of the
EGRET spectra. If the COMPTEL spectral points of the 4 VPs are
combined to the EGRET spectra for VP 19 (the only ones publicly available),
the 10-30~MeV flux points are again consistent with the extrapolations 
of the EGRET spectra, supporting our interpretation that 3C~454.3 and CTA~102
are the main emitting sources in Fig.~1.
The spectra show the trend of a flattening at MeV energies for both sources
(Fig.~2).

\subsection{Complete Mission}
\subsubsection{Detections}
No source is significantly detected (threshold 3$\sigma$) in this
sky region in individual VPs in any of the 
four standard (0.75-1, 1-3, 3-10, 10-30~MeV) COMPTEL bands.  
In order to improve statistics, we combined the observations listed in 
Table~1 into 2 periods: CGRO Phases 1 to 4 (simultaneous to the third EGRET catalog)
and the sum of all data, 
i.e. the complete COMPTEL coverage of this sky region during the 
CGRO mission.
This results in skymaps showing significant MeV emission above 3~MeV 
from this sky region. 

The 10-30 MeV maps (Fig.~3) show again emission peaks of 
$\sim$4.5$\sigma$ (Phases 1 to 4) and $\sim$3.9$\sigma$ (all mission), 
located in between 3C~454.3 and CTA~102, now with extensions towards
3EG~J2255+1943. Again, COMPTEL can not resolve the 5 EGRET sources, 
and therefore we check the EGRET measurements for this time period.
 Fig.~4 shows the measured EGRET spectra/fluxes of the 5 sources 
time-averaged for the CGRO Phases 1 to 4.
Because 3C~454.3 and CTA~102 are expected to be the brightest MeV
emitters, we naturally consider both of them as prime candidates
for this emission. For 3EG~J2243+1509 is only an upper limit given 
in the third EGRET catalog (only detected in VP~26). 
Therefore we neglect its contribution to the MeV excess, 
although it is the source closest to the emission peak. 
The contributions from 3EG~J2248+1745 and 3EG~J2255+1943
to this 10-30 MeV excess are expected to be at the same level, 
although a factor of 5 to 10 below the expected contributions of
3C~454.3 and CTA~102. Therefore we would not expect much contribution 
from these sources.
However, the maps of Fig.~3 suggest, by their extensions 
towards 3EG~J2255+1943, a significant contribution of this source, 
whose EGRET spectral index for the P1234 time period, 2.36$\pm$0.61,  
is rather uncertain (Hartman et al. 1999).
As suggested by the maps and the coincidence in time
(the extensions are consistent 
with the flaring of this source in VPs 336 and 410), we included 
3EG~J2255+1943 in the flux fitting procedure.  
Subsequently, the fluxes of three sources, 3C~454.3, CTA~102 and 3EG~J2255+1943,
are listed in Table~2 for the Phases 1-4 and Phases 1-7. They are obtained
by a simultaneous three-source fit.
We like to note that the 10-30 MeV fluxes of CTA~102 are not dependent 
on the inclusion of 3EG~2255+1943 in the fitting procedure, while the 
flux values of 3C 454.3 vary within a factor of 2 with, however, overlapping
error bars.   
      
In the 3-10 MeV band, a significant excess of $\sim$5.3$\sigma$ (one degree of
freedom) is detected in this sky region (Fig.~5) for the sum   
of all 14 VPs. Because only 3C~454.3 is located inside the 3\sig\-
error contour -- 4.5$\sigma$ detection significance at the
actual source position -- 
we attribute this emission solely to 3C~454.3.

\subsubsection{Flux variability}
A search for flux variability has been carried out for 3C~454.3 and CTA~102
in the four standard energy bands using all individual VPs.
All MeV light curves are statistically consistent with a constant flux
for both sources during the analyzed period of 7 years,
although flux variations were found in neighboring bands by
OSSE (McNaron-Brown et al. 1995)
and EGRET (Mukherjee et al. 1997) for both sources.
The COMPTEL 10-30~MeV detections are weak for both sources.
The trend is that COMPTEL detects the sources during the early
mission (3C~454.3 in Phase I; CTA~102 in Phase I and III) at periods
when EGRET reported flaring activity (Hartman et al. 1999). At later periods 
only non-detections are derived. In contrast to CTA~102 we find significant evidence
for 3C~454.3 in the sum of the COMPTEL 3-10~MeV data. Subdividing 
this emission into time bins according to the CGRO phases, 
indicates a weak (near the detection threshold of COMPTEL)
but likely stable MeV emission (Fig.~6). 
This is consistent with the significant detection in the sum of all data, 
and the non-detections in individual VPs. 
 
\subsubsection{Energy Spectra}
To further investigate the MeV properties of both blazars, we 
generated MeV spectra for different time periods. 
The time-averaged ones, containing all COMPTEL mission data, 
are shown in Fig.~7. The spectrum of 3C~454.3 indicates a power maximum
(at least with respect to MeV energies) in the 3-10 MeV band.
No conclusion can be drawn from the spectrum of CTA~102
because only the 10-30 MeV band yields a weak detection.
To combine the COMPTEL and EGRET spectra simultaneously (Fig.~8),
we generated time-averaged COMPTEL spectra for the CGRO Phases 1-4, covering the 
time period of the third EGRET catalog.
The COMPTEL data points are consistent with the EGRET spectral
extrapolations within the 1~sigma error limits in photon index;
though showing the trend of a spectral turnover.  
To check on this trend, we added non-simultaneous published
(McNaron-Brown et al. 1995) OSSE data, 
measured for both sources in 1994 (VPs 317, 319, 319.5 and 323 for
3C~454.3; VPs 316, 317 and 323 for CTA~102).
These combined OSSE/COMPTEL/EGRET spectra show now clearly
the spectral turnover with a power maximum in the COMPTEL range
for both sources (Fig.~9).
OSSE measured a photon index of 1.5 $^{+0.6}_{-0.4}$ for 3C~454.3
in VP~317, while EGRET reports a time-averaged spectral index of
2.21$\pm$0.06 (Hartman et al. 1999).
For CTA~102, OSSE observed a photon index of 1.0$^{+0.7}_{-0.6}$ in 
VP 323 (McNaron-Brown et al. 1995), while the third EGRET catalog reports
a time-averaged spectral index of 2.45$\pm$0.14. 
Therefore the changes in spectral index from high energy $\gamma$-rays
to hard X-rays have, at least for CTA~102, to be larger than 0.5.

In order to obtain a view on the general -- radio to \grays\ --
radiation behaviour of both blazars, we supplemented these
CGRO \gray\ data by published (non-simultaneous) measurements at
lower energies (Fig.~10). The two multifrequency spectra are generally similar,
showing both two spectral maxima. The low-energy ones are in the IR-band, 
while the high-energy ones, dominating the source luminosity, are at
MeV-energies.

\section{Discussion}
%######################################################################
Our analyses on the combined COMPTEL data from 1991 to 1998
show weak detections of 3C~454.3 and CTA~102 at energies above 10 MeV,
and a significant detection of 3C~454.3 in the 3-10 MeV band.
The time-averaged MeV spectra are consistent with the
extrapolations of the EGRET spectra (Fig.~8), while 
the addition of the OSSE hard X-ray data reveal for both sources
a significant spectral turnover with a power maximum at lower MeV energies (Fig.~9).
Assuming a Friedmann universe, H$_{0}$=75 km s$^{-1}$ Mpc$^{-1}$ and
q$_{0}$=0.5, we derive isotropic luminosities of 1.6$\times$10$^{48}$ erg s$^{-1}$
and 7.5$\times$10$^{47}$ erg s$^{-1}$  for 3C~454.3 (above 3 MeV)
and CTA~102 (above 10 MeV), respectively.
The corresponding minimum mass of the central black holes by assuming 
Eddington-limited accretion are 1.1$\times$10$^{10}$ M$_{\odot}$ (3C~454.3)
and 5.5$\times$10$^{9}$ M$_{\odot}$ (CTA~102) in the Thompson regime. 
These values lower to 2.85$\times$10$^{8}$ M$_{\odot}$ and 
8$\times$10$^{7}$ M$_{\odot}$ in the Klein-Nishina regime.
From OSSE variability measurements in 1995, McNaron-Brown et al. (1995)
derive upper mass limits for the central black holes of 
5.6$\times$10$^{10}$ M$_{\odot}$ for 3C~454.3 and
8.9$\times$10$^{10}$ M$_{\odot}$ for CTA~102.

Short-term variability together with large luminosities are
two main characteristics of EGRET-detected $\gamma$-ray blazars. 
Both facts together require a small photon-dense emission 
region. However, for several blazars -- among them 
3C~454.3 and CTA~102 (McNaron-Brown et al., 1995) --
the derived huge photon densities imply the 
pair production optical depth $\tau_{\gamma\gamma}$ 
to be larger than 1, which would lead to a cut-off of the \gray\ spectrum.
To avoid this contradiction, beaming of the photons is generally assumed. 
Therefore current models for the $\gamma$-ray emission in blazars
were developed in a beaming scenario, which generally assumes 
that the $\gamma$-ray emission is produced through inverse-Compton (IC)
scattering of photons off relativistic particles in a plasma jet
whose orientation is close to our line of sight.

The primary accelerated jet particles can be either electrons or protons.
Both  are successful in interpreting the spectral energy
distribution (SED) of blazars of different types, like 
flat-spectrum radio quasars (FSRQs) or BL~Lac objects.
3C~454.3 and CTA~102 belong to FSRQ-class of blazars. 
 A typical property of blazars is a two-hump spectrum 
with low-energy peaks from infrared to X-ray energies and high-energy
peaks from MeV up to TeV energies. Such spectra are usually
interpreted as synchrotron and inverse-Compton (IC) emission from
a relativistic jet.
In a leptonic scenario, the IC models can be categorized  
according to the origin of the soft photons
into synchrotron-self Compton (SSC) models (the self-generated 
synchrotron photons are the target photons of the relativistic jet leptons, 
e.g. Maraschi et al. 1992, Bloom $\&$ Marscher 1996) and 
so-called external Compton (EC) models (the soft target photons 
originate from outside the jet, e.g. Dermer $\&$ Schlickeiser 1993,
Sikora, Begelman $\&$ Rees 1994, Blandford $\&$ Levinson 
1995). 
In pure SSC models, the high-energy peak relates to the low-energy one 
by $\nu_{C}\propto\langle\gamma^{2}\rangle\nu_{S}$, where $\nu_{S}$ is
the frequency of the synchrotron peak,  $\nu_{C}$ the frequency of 
the SSC peak, and $\langle\gamma\rangle$ the average Lorentz-factor
(energy) of the electrons.
In a SSC scenario, $\gamma$ should be in a range of 10$^{3}$
 - 10$^{5}$ for 3C~454.3 and CTA~102 according to the peak positions
in their SEDs (Fig.~10). The EC process is always taken into account
in modeling the $\gamma$-ray flaring behaviour of blazars,
due to its stronger dependence on the bulk Lorentz factor. The SEDs of the
flaring $\gamma$-ray blazars PKS 0528+134, 3C 273 and 3C 279 are well 
interpreted in such a multi-component scenario (e.g. Mukherjee et al. 1999,
Collmar et al. 2000, Hartman et al. 2001). However, according to the 
investigations on PKS~0528+134 by Mukherjee et al. (1999) and on 
3C~279 by Hartman et al. (2001), the contribution of the EC process is
also important, if the blazar is in a low  $\gamma$-ray state.
Furthermore, Sambruna et al. (1997) found that the EC scattering may
be the primary cooling mechanism  in the low $\gamma$-ray state
of PKS~0528+134, if the $\gamma$-ray flux exceeds the optical one.

Ghisellini et al. (1998) modeled the SEDs of 51 $\gamma$-ray blazars
with similar SEDs as in Fig.~10 (among them 3C~454.3 and CTA~102)  
by two types of models: a pure SSC model and a combined 
SSC/EC model. They assumed a black body distribution for the soft
EC photons, which are radiated from an optically thick accretion disk. 
They found that for most blazars the SED can in principle be
represented by both types of models. They suggest that the sequence  
{\it high frequency peaked BL Lac object} $\rightarrow$
 {\it low frequency peaked BL Lac object}
$\rightarrow$ {\it flat-spectrum radio quasars} 
(HBL$\rightarrow$LBL$\rightarrow$FSRQs) may be related to an 
increasing dominance of the EC over the SSC emission.
Ghisellini et al. (1998) note that the bulk Lorentz factor, $\Gamma$, has
to be of the order of 10-15 to produce a  power maximum 
peaking at  MeV energies. 
In EC models the IC target photons  can be the direct accretions disk photons
(ECD; e.g. Dermer, Schlickeiser $\&$ Mastichiadis 1992, Dermer $\&$ Schlickeiser 1993), 
the disk photons reflected by the broad-line region 
(ECC; e.g. Sikora, Begelman $\&$ Rees 1994) and the synchrotron photons 
reflected by the broad-line region (Rsy; e.g., Ghisellini $\&$ Madau 1996). 
In the modeling of PKS~0528+134 (Mukherjee et al. 1999) and 3C~279  
(Hartman et al. 2001) a bulk Lorentz factor strength between $\sim$5 and $\sim$20 
is found, which leads to a narrow ECD emission component sticking out at MeV 
energies on top of a broad SSC component.

The most interesting new result is the evidence for a longterm steady 
emission of 3C~454.3 in the COMPTEL 3-10~MeV band, which -- at least on time 
average -- covers the IC peak of the blazar emission.
Interpreting our data in the framework of leptonic multicomponent models,
we attribute this MeV emission to an ECD component on top of an 
SSC component, given the high bulk Lorentz factor derived by Ghisellini
et al. (1998) for 3C~454.3.
This rather stable MeV emission does not agree with the time variability 
observed by EGRET at energies above 100~MeV (Fig.~11).
While at EGRET energies the flux level decreased by a factor of $\sim$3
(Hartman et al. 1999), the COMPTEL 3-10~MeV flux level stays the same.
These different variability behaviours probably suggest that different 
emission processes are at work in these two energy bands.
Such a behaviour is consistent with the multicomponent modeling of 3C~279 
by Hartman et al. (2001). In this framework, we would attribute 
the EGRET part of the spectrum to the ECC component, while, as already 
mentioned above, the COMPTEL part to the ECD component. 
In any case, between Phase~1 and Phase~3 something has changed 
inside 3C~454.3, which leaves the MeV emission unaffected, while 
significantly changes the flux at energies above 100~MeV.
Due to the different dependence of the ECD and ECC on the bulk 
Lorentz factor $\Gamma$ (B\"ottcher 1999), this peculiar variability 
behaviour could reflect a reduction in $\Gamma$ from Phase~1 to Phase~3.    

This uncorrelated variability behaviour is similar to the one 
observed in 3C~273 (Collmar et al. 2000), where during a two-week 
flaring event at EGRET energies, the MeV-flux at COMPTEL energies 
did not follow this high-energy flare. Apart from the different time-scales, 
these two observations are quite similar. Also in other respects, 
3C~454.3 has similar MeV properties as 3C~273. Both blazars 
seem to have a rather stable MeV emission over years, and a clear 
spectral maximum in the 3-10~MeV band (Collmar et al. 2001). This
probably suggests that the same physical processes are responsible
for their MeV radiation.

\section{Conclusion}
%######################################################################

We report the results of an analysis of all COMPTEL data of 
the \gray\ blazars 3C~454.3 and CTA~102. 
Our results of the early mission (Phase 1) 
are consistent with the ones of Blom et al. (1995). We conclude that 
the observed excess in the 10-30 MeV band during this early 
period, located in between the 
two blazars, is dominated by 3C~454.3 and CTA~102, but might have 
some minor contributions from nearby unidentified EGRET
$\gamma$-ray sources.
In the time-averaged (Phase 1 to 4 and Phase 1 to 7) data,
we obtain again significant MeV emission in the 
uppermost (10-30~MeV) COMPTEL band, which we attribute to 3C~454.3, 
CTA~102, and 3EG~J2255+1943.  
We find a strong detection of 3C~454.3 in the 3-10~MeV band.
Time-resolved analyses suggest a likely steady
(although near the COMPTEL threshold) MeV emission
from 3C~454.3 in the 3-10~MeV band, which yields the strong detection 
in the sum of all COMPTEL data.   
This likely steady MeV emission is different to the significantly observed
time variability at EGRET energies. Given this fact, we conclude that 
different emission mechanisms dominate at these energy bands. 
According to current leptonic multicomponent emission scenarios,  
this may be the Comptonization of direct accretion disk photons 
at COMPTEL energies, and at EGRET energies the Comptonization of
accretion disk photons which have been scattered into the jet region 
by the broad-line region clouds.

The MeV spectra of both sources are consistent with the spectral
extrapolations of the contemporaneous EGRET spectra. However, the inclusion of 
(non-simultaneous) OSSE data reveals a strong spectral turnover at MeV energies.
Putting the COMPTEL data into multifrequency perspective, reveals that 
-- at least on time average -- the peak of the IC emission, which dominates the 
luminosity, is for both sources at MeV energies.

\acknowledgements
This research was supported by the German government through 
DLR grant 50 QV 9096 8, by NASA under contract NAS5-26645, and by the Netherlands
Organization for Scientific Research NWO. S. Zhang was also subsidized by the 
Special Funds for Major State Basic Research Projects
and by the National Natural Science Foundation of China under grant 10373013.

\newpage

%%%%% Fig 1 %%%%%%%%%%%%%%%%%%%%%%%%%%%%%%%%%%%%%%%%%%%%
\begin{figure}[t]
\centerline{
  \epsfig{figure=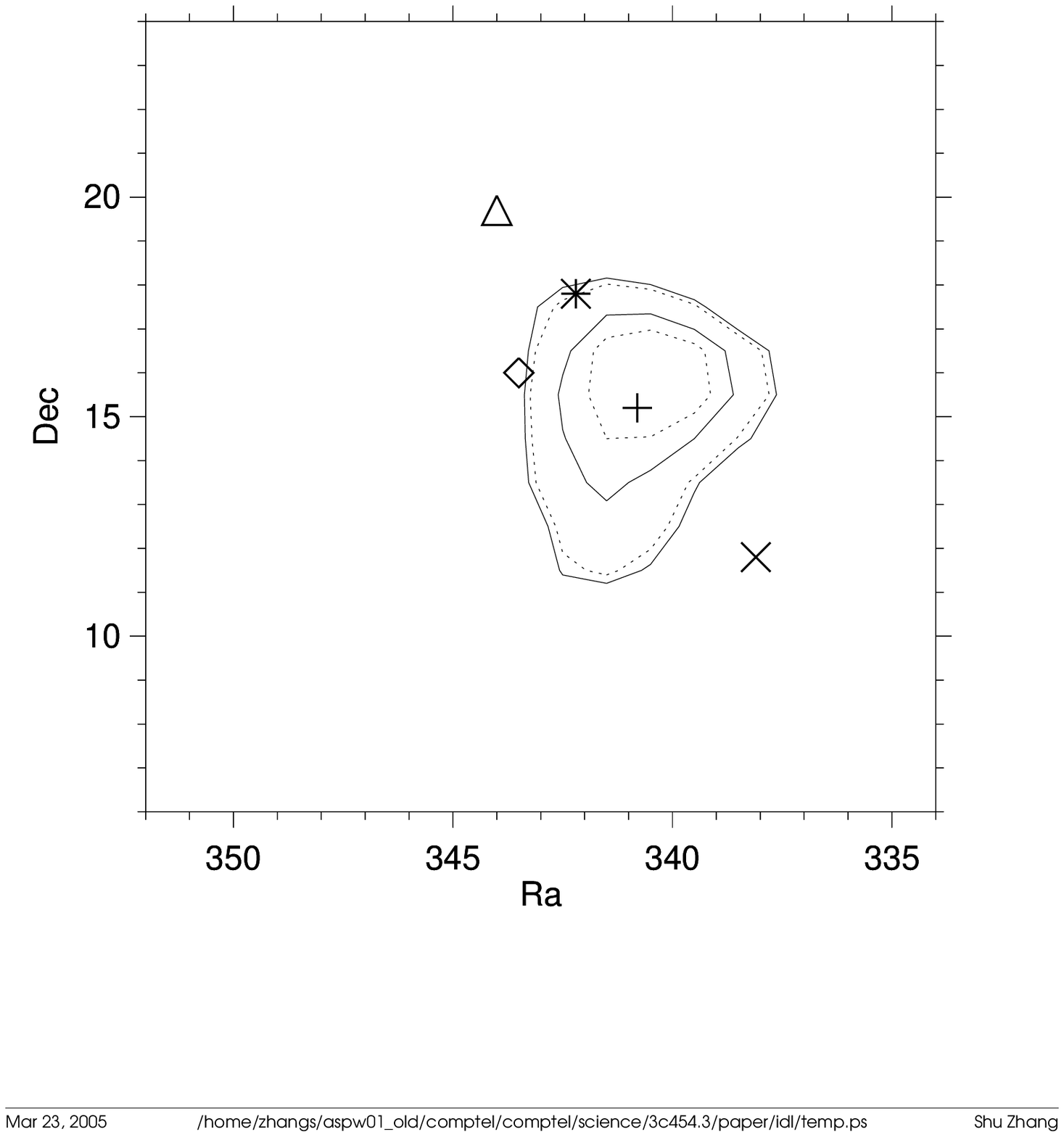,height=7.0cm,width=7.0cm,clip=} \hfill
  \epsfig{figure=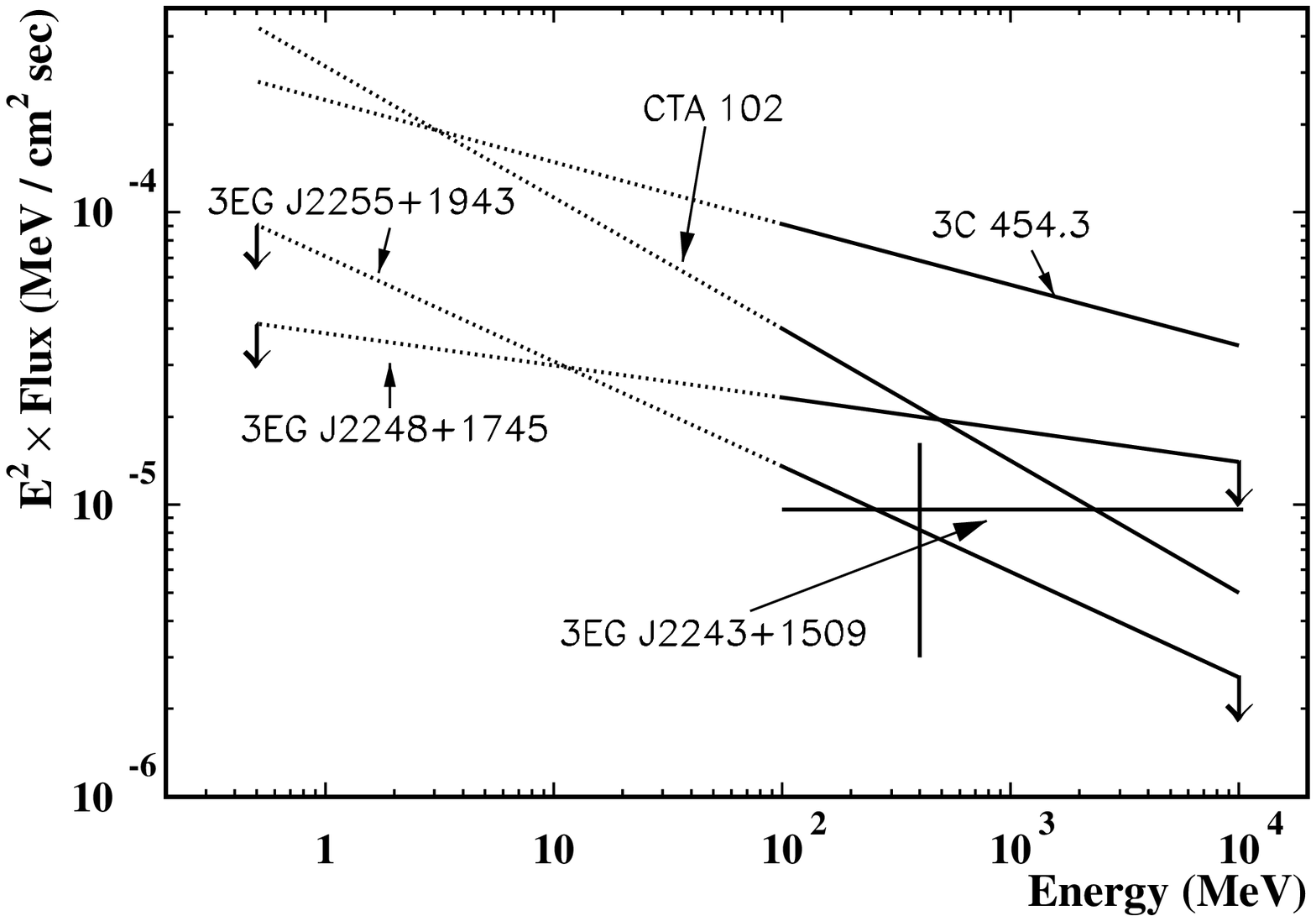,width=7.0cm,clip=}
}
\caption{
Left: COMPTEL 10-30~MeV significance (solid contours) and error location
(dotted contours) map of the 3C~454.3 ($\diamond$)
and CTA~102 ($\times$) sky region
for CGRO Phase 1 (May 1991 to November 1992).
The locations of three nearby EGRET sources 3EG~J2243+1509 (+), 
3EG~J2248+1745 ($\ast$),
and 3EG~J2255+1943 ($\triangle$) in this sky region are overlaid.
The significance contour lines start at 3$\sigma$ 
(outer solid contour) with steps of 0.5$\sigma$,
and the error location contour lines at 1$\sigma$ 
(inner dotted contour) with steps of 1$\sigma$.
\newline
Right:  
Best-fit spectral shapes in the EGRET band (solid lines) 
and their spectral extrapolations throughout the COMPTEL band (dotted lines) 
are shown for 4 of the 5 EGRET sources of the map, according to the source 
parameters given in  the third EGRET catalog (Hartman et al. 1999) for 
CGRO Phase 1 (P1).
These shapes are generated by using the simultaneously (P1) measured
flux values and the given (non-simultaneous) spectral indices for P1234.
Note that the spectral shapes for 3EG~J2255+1943 and 3EG~J2248+1745
are upper limits.
For 3EG~J2243+1509 the third EGRET catalog does not provide
a spectral shape. In order to estimate its contribution,  
we plot its simultaneous flux value at an assumed center of
energy of 400~MeV. 
3C~454.3 and CTA~102 should be the strongest MeV sources.}
\end{figure}
%%%%% Fig 1 %%%%%%%%%%%%%%%%%%%%%%%%%%%%%%%%%%%%%%%%%%%%

%%%%% Fig 2 %%%%%%%%%%%%%%%%%%%%%%%%%%%%%%%%%%%%%%%%%%%%
\begin{figure}[b]
\centering
\psfig{figure=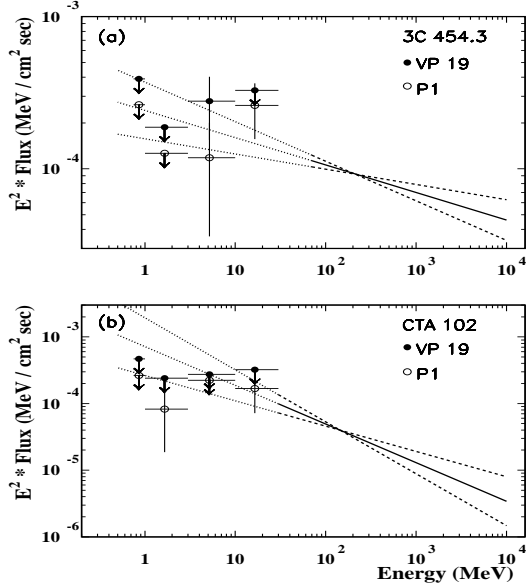,height=8.0cm,width=8.0cm,clip=}
\caption{Combined COMPTEL/EGRET spectra of 3C~454.3 (a) and CTA~102 (b).
The COMPTEL data are from complete CGRO Phase 1 (open circles) and VP 19 
(filled circles). The error bars are 1$\sigma$
and the upper limits are 2$\sigma$. The EGRET spectra are from VP 19.
The solid lines represent the best-fit EGRET spectra, the dashed
lines their 1-$\sigma$ error limits in slope, and the dotted lines
their extrapolations towards MeV energies.}
\end{figure}
%%%%% Fig 2 %%%%%%%%%%%%%%%%%%%%%%%%%%%%%%%%%%%%%%%%%%%%

%%%%% Fig 3 %%%%%%%%%%%%%%%%%%%%%%%%%%%%%%%%%%%%%%%%%%%%
\begin{figure}[t]
\centerline{
  \epsfig{figure=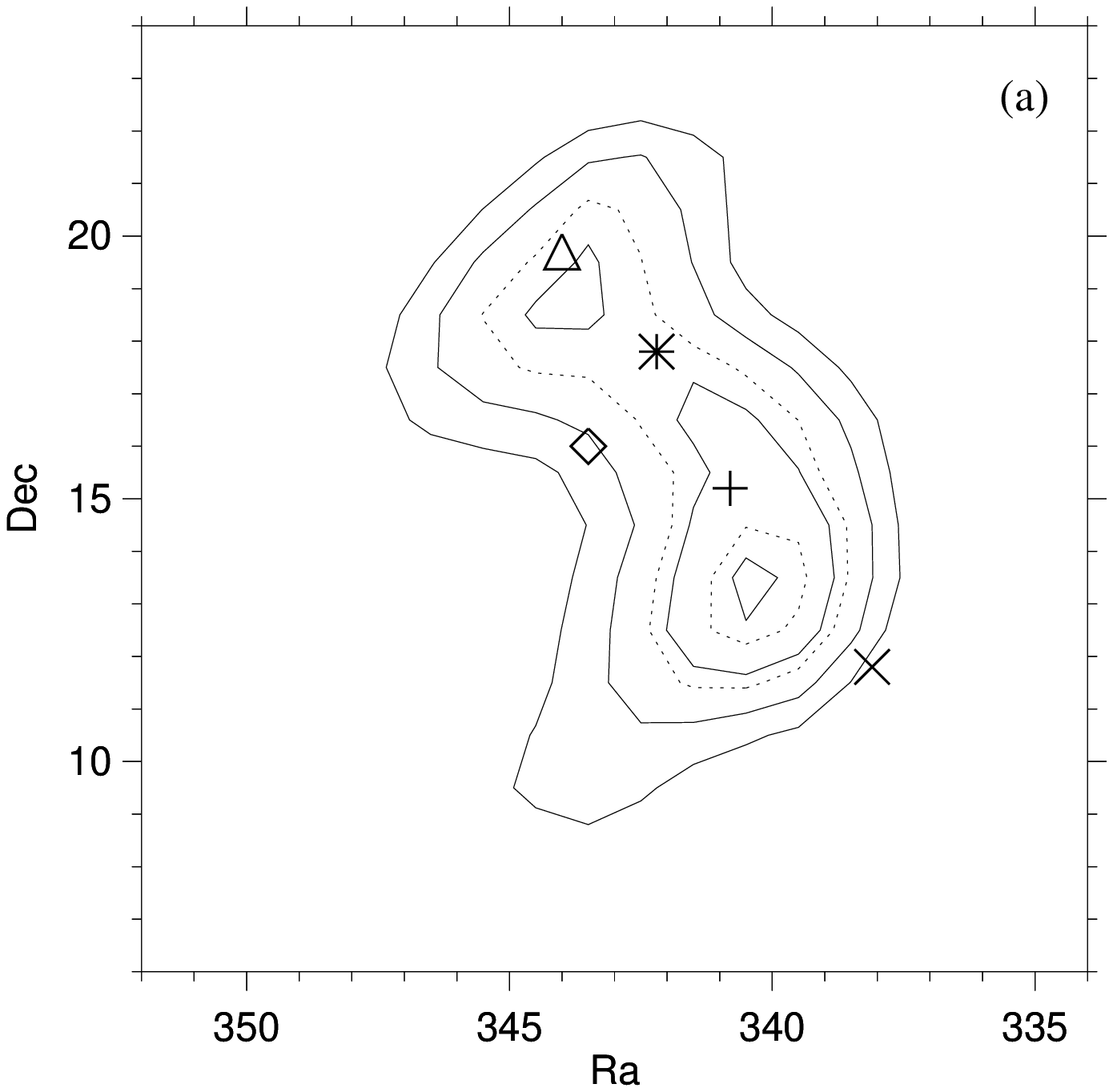,width=7.0cm,clip=} 
  \epsfig{figure=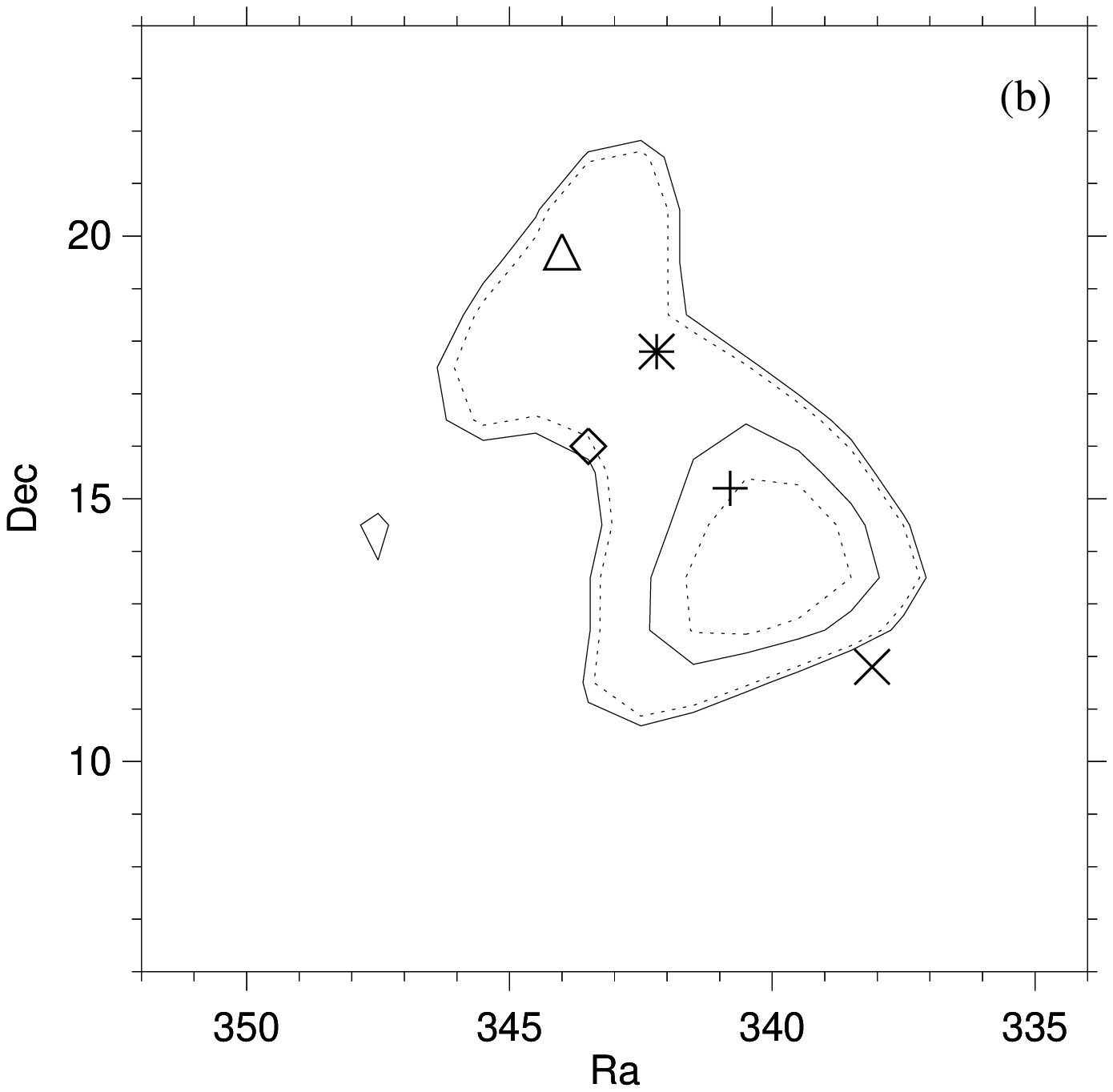,width=7.0cm,clip=}
}
\caption{
 COMPTEL 10-30 MeV significance (solid contours) and error location
(dotted contours) map of the 3C~454.3 ($\diamond$)
and CTA~102 ($\times$) sky region for a) CGRO Phase 1 to 4
(corresponding to P1234 in the third EGRET catalog),
and b) the sum of all COMPTEL data.
The locations of three nearby EGRET sources 3EG~J2243+1509 (+),
3EG~J2248+1745 ($\ast$),
and 3EG~J2255+1943 ($\triangle$) in this sky region are overlaid.
The significance contour lines start at 3$\sigma$ 
(outer solid contour) with steps of 0.5$\sigma$,
and the error location contour lines at 1$\sigma$ 
(inner dotted contour) with steps of 1$\sigma$.}
\end{figure}
%%%%% Fig 3 %%%%%%%%%%%%%%%%%%%%%%%%%%%%%%%%%%%%%%%%%%%%

%%%%% Fig 4 %%%%%%%%%%%%%%%%%%%%%%%%%%%%%%%%%%%%%%%%%%%%
\begin{figure}[b]
\centerline{\epsfig{figure=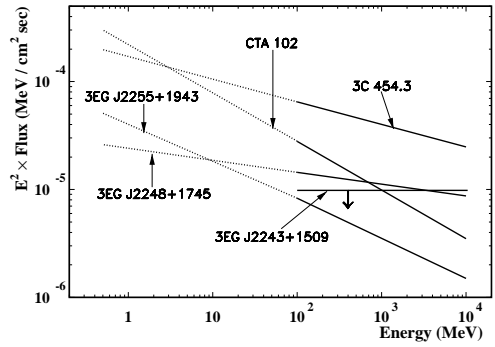,width=7.0cm,clip=}} 
\caption{
 Best-fit spectral shapes in the EGRET band (solid lines) 
and their spectral extrapolations throughout the COMPTEL band (dotted lines) 
are shown for 4 of the 5 EGRET sources of this sky region according to the source 
parameters given in  the third EGRET catalog (Hartman et al. 1999).  
These shapes are generated by using the P1234 measured flux values
and spectral indices.
For 3EG~J2243+1509 the third EGRET catalog does not provide
a spectral shape. In order to estimate its contribution,  
we plot its P1234 upper flux limit at an assumed center of
energy of 400~MeV. 
3C~454.3 and CTA~102 should be the strongest MeV sources.}  
\end{figure}
%%%%% Fig 4 %%%%%%%%%%%%%%%%%%%%%%%%%%%%%%%%%%%%%%%%%%%%

%%%%% Fig 5 %%%%%%%%%%%%%%%%%%%%%%%%%%%%%%%%%%%%%%%%%%%%
\begin{figure}[t]
\centering
\epsfig{figure=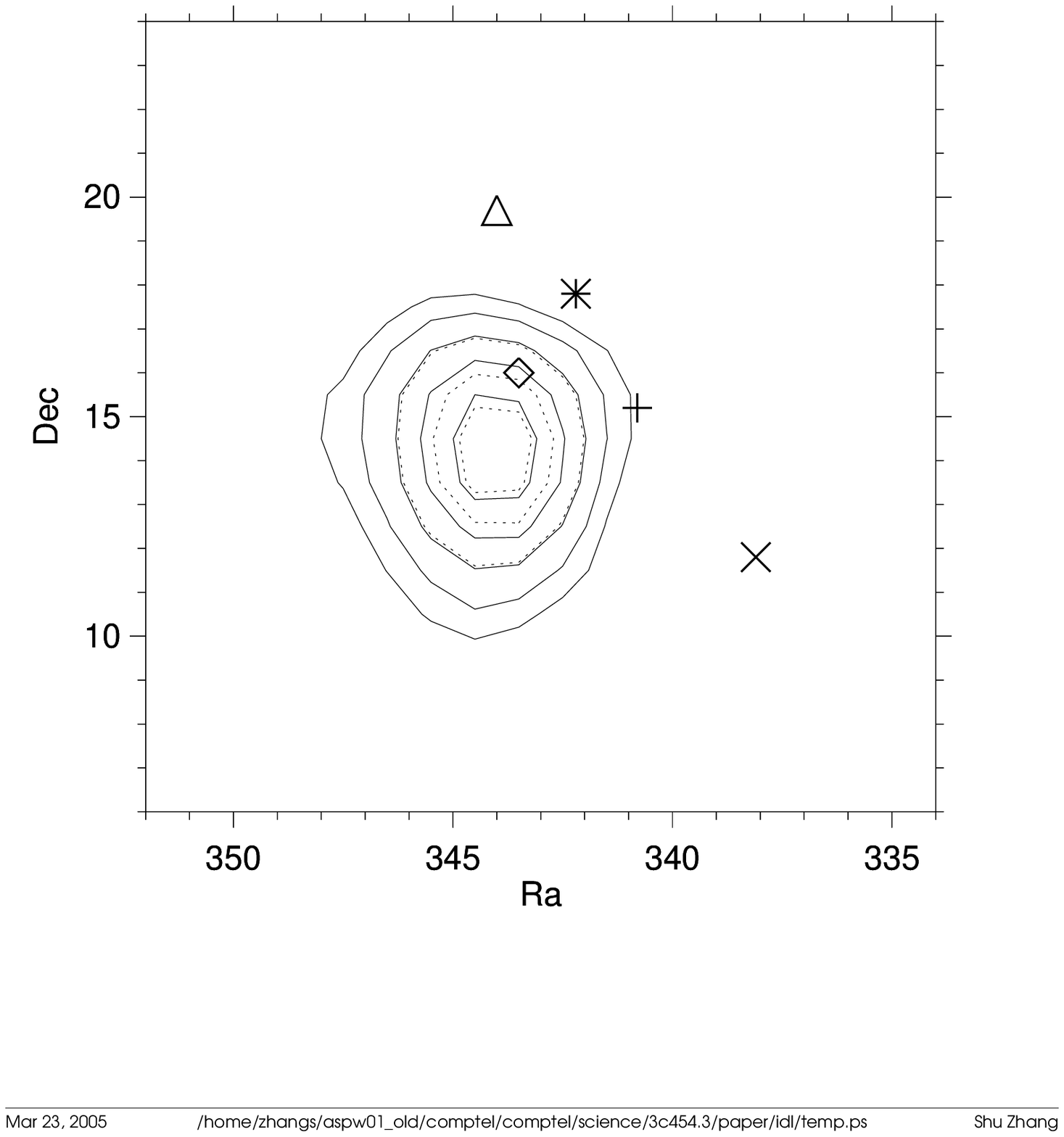,height=8.0cm,width=8.0cm,clip=}
\caption{COMPTEL 3-10 MeV significance (solid contours) and error location
(dotted contours) map of the 3C~454.3 ($\diamond$)
and CTA~102 ($\times$) sky region for the sum of all COMPTEL data.
The locations of three other nearby EGRET sources, 3EG J2243+1509 (+),
3EG J2248+1745 ($\ast$), and 3EG J2255+1943 ($\triangle$), are overlaid.
 The significance contour lines start at 3$\sigma$ 
(outer solid contour) with steps of 0.5$\sigma$,
and the error location contour lines at 1$\sigma$ 
(inner dotted contour) with steps of 1$\sigma$.}
\end{figure}
%%%%% Fig 5 %%%%%%%%%%%%%%%%%%%%%%%%%%%%%%%%%%%%%%%%%%%%

%%%%% Fig 6 %%%%%%%%%%%%%%%%%%%%%%%%%%%%%%%%%%%%%%%%%%%%
\begin{figure}[b]
\centering
\epsfig{figure=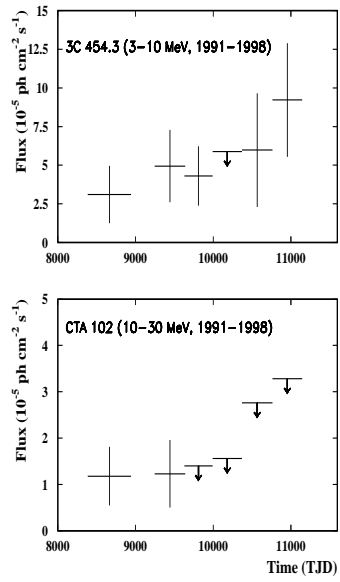,height=8.0cm,width=6.0cm,clip=}
\caption{The 3-10 MeV light curve of 3C~454.3 (upper panel)
and the 10-30 MeV light curve of CTA~102 (lower panel) 
with  each bin averaged over one CGRO Phase.
Phase 2 is missing due to the lack of observations in this period.
The error bars are 1$\sigma$ and the upper limits are 2$\sigma$.}
\end{figure}
%%%%% Fig 6 %%%%%%%%%%%%%%%%%%%%%%%%%%%%%%%%%%%%%%%%%%%%

%%%%% Fig 7 %%%%%%%%%%%%%%%%%%%%%%%%%%%%%%%%%%%%%%%%%%%%
\begin{figure}[t]
\centering
\epsfig{figure=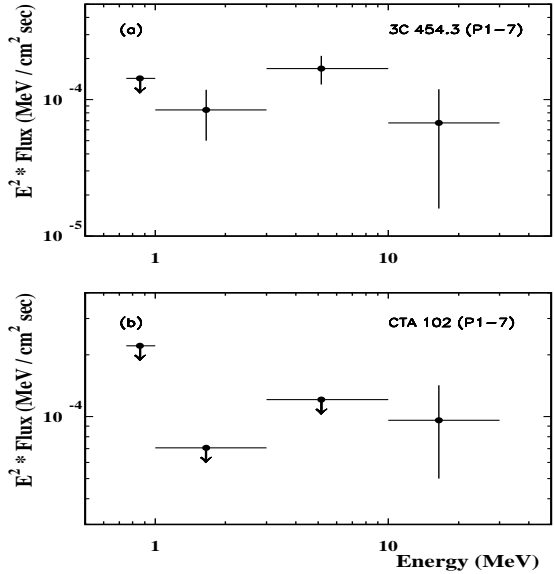,height=8.0cm,width=10.0cm,clip=}
\caption{The COMPTEL spectra of 3C~454.3 (a) and CTA~102 (b) for the 
sum of all COMPTEL data. The error bars are 1$\sigma$ and the upper
limits are 2$\sigma$.}
\end{figure}
%%%%% Fig 7 %%%%%%%%%%%%%%%%%%%%%%%%%%%%%%%%%%%%%%%%%%%%

%%%%% Fig 8 %%%%%%%%%%%%%%%%%%%%%%%%%%%%%%%%%%%%%%%%%%%%
\begin{figure}[b]
\centering
\epsfig{figure=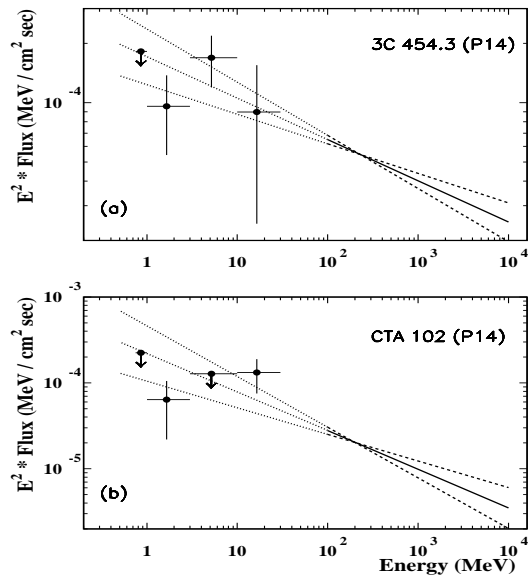,height=8.0cm,width=8.0cm,clip=}
\caption{
Simultaneous COMPTEL and EGRET spectra of 3C~454.3 (a) and CTA~102 (b) for the CGRO
Phases 1 to 4. The filled circles represent the COMPTEL spectral points.
The solid lines represent the best-fit EGRET spectra, 
the dashed lines their 1-$\sigma$ error limits in slope,
and the dotted lines the extrapolations into the COMPTEL band.
}
\end{figure}
%%%%% Fig 8 %%%%%%%%%%%%%%%%%%%%%%%%%%%%%%%%%%%%%%%%%%%%

%%%%% Fig 9 %%%%%%%%%%%%%%%%%%%%%%%%%%%%%%%%%%%%%%%%%%%%
\begin{figure}[tbh]
\centering
\epsfig{figure=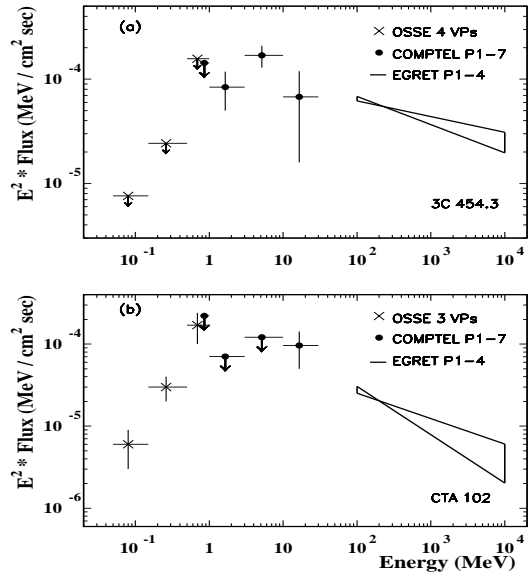,height=8.0cm,width=8.0cm,clip=}
\caption{
Non-Simultaneous OSSE, COMPTEL and EGRET \gray\ spectra of 3C~454.3 (a)
and CTA~102 (b).
The solid lines represent the EGRET spectra inside the 1$\sigma$
error limits in slope. The OSSE results (McNaron-Brown et al. 1995) 
are from 4 VPs in 1994 for 3C~454.3, and from 3 VPs in 1994 for CTA~102
(see text). The EGRET spectra (Hartman et al. 1999) are from the first
4 CGRO Phases, i.e. time-averaged between April 1991 and November 1994 .
}
\end{figure}
%%%%% Fig 9 %%%%%%%%%%%%%%%%%%%%%%%%%%%%%%%%%%%%%%%%%%%%

%%%%% Fig 10 %%%%%%%%%%%%%%%%%%%%%%%%%%%%%%%%%%%%%%%%%%%%
\begin{figure}[tbh]
\centering
\epsfig{figure=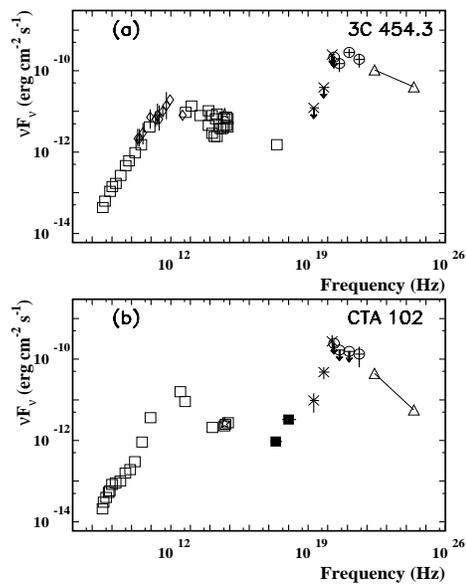,height=8.0cm,width=8.0cm,clip=}
\caption{Non-simultaneous SEDs of 3C~454.3 (a) and CTA~102 (b).  
The solid lines with triangles on each end represent the 
time-averaged (Phase 1-4) EGRET spectra (Hartman et al. 1999),
open circles the COMPTEL measurements (all mission), '$\ast$' the OSSE 
observations in 1994 (McNaron-Brown et al. 1995), and filled squares the ASCA
measurements in 1995 (taken from NED). The low-energy data symbolized
by open diamonds are taken from Montigny et al. (1995), and the ones 
symbolized by open squares from Impey et al. (1988).}
\end{figure} 
%%%%% Fig 10 %%%%%%%%%%%%%%%%%%%%%%%%%%%%%%%%%%%%%%%%%%%%

%%%%% Fig 11 %%%%%%%%%%%%%%%%%%%%%%%%%%%%%%%%%%%%%%%%%%%%
\begin{figure}[tbh]
\centering
\epsfig{figure=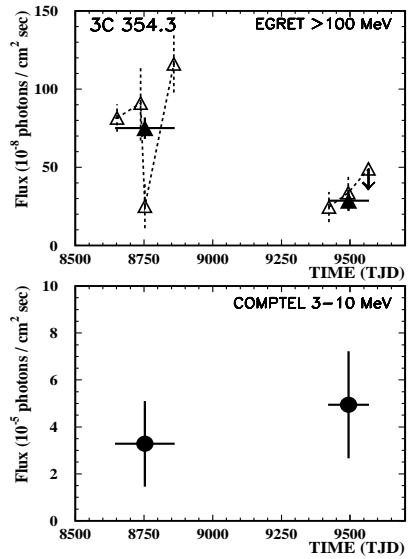,height=8.0cm,width=6.0cm,clip=}
\caption{Simultaneous EGRET ($>$100~MeV; $\triangle$) and COMPTEL (3-10~MeV; circles)
flux measurements of 3C~454.3. The filled triangles represent the average
EGRET fluxes of CGRO Phase 1 and 3 (Hartman et al. 1999),
while the open triangles the fluxes as measured
in individual VPs. The obvious and significant flux decline (factor of $\sim$3)
at energies above 100~MeV from Phase 1 (in 1992) to Phase 3 (in 1994) has no
counterpart at COMPTEL energies, indicating an independent variability behaviour
at the two bands. The error bars are 1$\sigma$.   
 }
\end{figure} 
%%%%% Fig 11 %%%%%%%%%%%%%%%%%%%%%%%%%%%%%%%%%%%%%%%%%%%%

%%%%% Table 1 %%%%%%%%%%%
\begin{table*}[hb]
\caption{COMPTEL observational periods of 3C~454.3 and CTA~102 during the whole  
CGRO mission, where at least one of the two objects was within 30\deg\ of
the pointing direction. The CGRO VPs, their time periods,
prime observational targets, offset angles, effective exposures and
the CGRO Phases  are given.}
\begin{flushleft}
\begin{tabular}{cccccccc}
\hline 
\multicolumn{1}{c}{VP}&\multicolumn{1}{c}{Date}&\multicolumn{1}{c}{ Target}&\multicolumn{2}{c}{Offset angle}&\multicolumn{2}{c}{Effective exposure (days)}&\multicolumn{1}{c}{CGRO Phase}\\  
\multicolumn{1}{c}{$\#$}&\multicolumn{1}{c}{ }&\multicolumn{1}{c}{ }&\multicolumn{1}{c}{3C~454.3}&\multicolumn{1}{c}{CTA~102}&\multicolumn{1}{c}{3C~454.3}&\multicolumn{1}{c}{CTA~102}&\multicolumn{1}{c}{}\\ \hline
19.0 & 23/01/92-06/02/92 &  G 058-43 & 15$^{\circ}$& 22$^{\circ}$&3.08&3.79 &Phase I\\ 
26.0 & 23/04/92-28/04/92 & MRK 335 & 18$^{\circ}$& 24$^{\circ}$  & 0.61&0.49 &\\ 
28.0 & 07/05/92-14/05/92 & MRK 335 & 18$^{\circ}$& 24$^{\circ}$  & 1.13&0.90 &\\ 
37.0 & 20/08/92-27/08/92 & MRK 335 & 15$^{\circ}$& 21$^{\circ}$  & 1.23&1.01 &\\ \hline
320.0 & 08/03/94-15/03/94& NGC 7469 & 8$^{\circ}$& 8$^{\circ}$  & 2.16&2.14 &Phase III\\  
327.0 & 17/05/94-24/05/94 &  Gal 083-50 & 12$^{\circ}$& 12$^{\circ}$  & 1.54&1.54 &\\  
336.0 & 01/08/94-04/08/94 & Gal 088-47 & 9$^{\circ}$& 11$^{\circ}$  & 0.96&0.90 &\\ \hline
410.0 & 24/01/95-14/02/95 & Gal 82-33 & 6$^{\circ}$& 7$^{\circ}$  & 4.97&4.91 &Phase IV\\ 
429.5 & 27/09/95-03/10/95& GRO J2058+42 & 26$^{\circ}$& 27$^{\circ}$  & 1.04&0.98 &\\  \hline
507.0 & 28/11/95-07/12/95 &  CTA~102 & 7$^{\circ}$& 0$^{\circ}$  & 2.16&2.49 &Phase V\\  
507.5 & 07/12/95-14/12/95 &  CTA~102 & 7$^{\circ}$& 0$^{\circ}$  & 1.68&1.94 &\\  
514.0 & 13/02/96-20/02/96 & Gal 60-60 & 27$^{\circ}$& 24$^{\circ}$  & 1.08&1.22 &\\ \hline
601.1 & 15/10/96-29/10/96& PSR J2043+2740& 31$^{\circ}$& 28$^{\circ}$  & 1.88&2.12 &Phase VI\\  
623.5 & 15/07/97-22/07/97 & Bl Lac & 28$^{\circ}$& 31$^{\circ}$  &1.06&0.92  &\\ \hline
713.0 & 24/02/98-10/03/98 & Gal 110-20& 24$^{\circ}$ & 30$^{\circ}$  & 2.61&1.97 &Phase VII\\ 
\hline
\end{tabular}\end{flushleft}
\label{tab1}
\end{table*}
%%%%% Table 1 %%%%%%%%%%%

%%%%% Table 2 %%%%%%%%%%%
\begin{table*}[htb]
\caption{Fluxes and upper limits of 3C~454.3, CTA~102 
and the unidentified EGRET source 3EG~J2255+1943 for different
observational periods in units of 10$^{-5}$ ph cm$^{-2}$ s$^{-1}$.
The errors are 1$\sigma$ and the upper limits are 2$\sigma$. An upper
limit is given when the significance of an individual flux value
is less than 1$\sigma$.} 
\begin{flushleft}
\begin{tabular}{ccccccccc}
\hline\noalign{\smallskip}
\multicolumn{1}{c}{Energy } & \multicolumn{3}{c}{3C~454.3} & \multicolumn{3}{c}{CTA~102}& \multicolumn{2}{c}{3EG J2255+1943} \\
 (MeV) & Phase 1 & Phases 1-4 & Phases 1-7 & Phase 1 & Phases 1-4 & Phases 1-7& Phases 1-4 & Phases 1-7   \\
\hline\noalign{\smallskip}
0.75-1 & $<$8.35       & $<$6.05       & $<$4.79        & $<$8.48       &  $<$7.56      & $<$7.40 & $<$ 7.6 & $<$ 6.1\\
1-3    & $<$8.58       & 6.40$\pm$2.78 & 5.59$\pm$2.30  & 5.03$\pm$4.03 & 4.26$\pm$2.80 & $<$4.64  & $<$6.73 & $<$ 5.49\\
3-10   & 2.96$\pm$1.85 & 3.94$\pm$1.16 & 3.95$\pm$0.95  & $<$5.26       & $<$2.92       & $<$2.83 & $<$2.73 & $<$2.2\\
10-30  & 1.74$\pm$0.69 & 0.60$\pm$0.43 & 0.45$\pm$0.34  & 1.12$\pm$ 0.64& 0.88$\pm$0.38 & 0.64$\pm$0.31& 1.03$\pm$0.41 &0.70$\pm$0.34 \\
 \hline\noalign{\smallskip}
\end{tabular}\end{flushleft}
\label{tab2}
\end{table*}
%%%%% Table 2 %%%%%%%%%%%

\end{document}